# Equivalence of pseudogap and pairing energy in a cuprate high-temperature superconductor

Jiasen Niu[1,2,7]†, Maialen Ortego Larrazabal[3]†, Thomas Gozlinski[1,2,7], Yudai Sato[1,2,7], Koen M. Bastiaans[4], Tjerk Benschop[1], Jian-Feng Ge[1,5], Yaroslav M. Blanter[4], Genda Gu[6], Ingmar Swart[3], Milan P. Allan[1,2,7],* .

[1] Leiden Institute of Physics, Leiden University, Niels Bohrweg 2, 2333 CA Leiden, The Netherlands.
[2] Fakultät für Physik, Ludwig-Maximilians-Universität, Schellingstrasse 4, München 80799, Germany.
[3] Debye Institute of Nanomaterials Science, Utrecht University, PO Box 80000, 3508 TA Utrecht, The Netherlands.
[4] Kavli Institute of Nanoscience, Delft University of Technology, 2628 CJ Delft, The Netherlands.
[5] Max Planck Institute for Chemical Physics of Solids, 01187 Dresden, Germany.
[6] Condensed Matter Physics and Materials Science Department, Brookhaven National Laboratory, Upton, NY 11973,USA.
[7] Munich Center for Quantum Science and Technology (MCQST), München, Germany

† These authors contributed equally.
* Corresponding author: milan.allan@lmu.de

**The pseudogap stands out in the phase diagram of the cuprate high-temperature superconductors because its origin and relationship to superconductivity remain elusive. The origin of the pseudogap has been debated, with competing hypotheses attributing it to preformed electron pairs or local order, such as charge density waves. Here, we present unambiguous evidence supporting the pairing scenario, using local shot-noise spectroscopy measurements in $Bi_2Sr_2CaCu_2O_{8+\delta}$. Our data demonstrates that the pseudogap energy coincides with the onset of electron pairing, and is spatially heterogeneous with values reaching up to 70 meV. Our results exclude a pure local order origin of the pseudogap, link the pseudogap to Cooper pair formation, and show that the limiting factor for higher $T_c$ in cuprates is phase coherence.**

In cuprate high-temperature superconductors, the Mott insulator and superconductivity phases are connected by the pseudogap phase(*1–4*). The pseudogap phase is bound by a crossover temperature $T_{PG}$, which decreases with doping; in contrast, superconductivity, which is bound by $T_c$, has a dome shape. These temperatures are reflected in two distinct spectroscopic features in tunneling experiments: a large gap in the spectral weight around the Fermi level (the so-called pseudogap), and a smaller gap-like signature that is often referred to as "kink"(*5–8*) (Fig. 1A, B). Intriguingly, from a spectroscopy point of view the pseudogap phase smoothly transitions into the superconducting phase below $T_c$, hinting that the mysterious pseudogap phase may hold the key to the high-$T_c$ puzzle. The origin of the pseudogap and its relation to superconductivity and the Mott state are among the most discussed – and controversial – open questions in quantum materials research; it has been a focus of experimental (*1, 2, 9*), theoretical (*3, 4, 10, 11*), and computational materials physics (*12*) as well as cold-atom physics (*13, 14*). The pseudogap energy scale $\Delta_{PG}$ is observed in tunneling experiments as a large, d-wave-like gap with energy $\Delta_{PG}$, (Fig. 1A), in photoemission as a gap of similar behavior at the antinode, and in transport as an increase in resistivity. Intriguingly, $\Delta_{PG}$ continues to increase at lower doping, even when superconductivity disappears. This again reflects the doping dependence of $T_{PG}$ (Fig. 1A), although we note that defining $T_{PG}$ accurately is challenging.

The plethora of theories describing the origin of the pseudogap phase can roughly be divided into two categories: one related to local orders and one related to preformed pairs (*1, 2, 15*). The local order picture is as follows: one assumes that local order – say a charge density wave – forms below $T_{PG}$, which leads to a depletion of spectral weight within $\Delta_{PG}$. In this scenario, superconductivity emerges independently from the local ordering at $T_c < T_{PG}$. In the superconducting state, the spectra will show two gaps: a large partial gap from the charge order and a smaller gap associated with superconductivity. This might be related to the kink-like signature (*5–8*) at the energy $\Delta_{kink}$ seen in tunneling experiments (Fig. 1B). Indeed, a multitude of local orders have been detected in the cuprates (*1, 2, 11*), some of which exhibit the same temperature dependencies as $T_{PG}$ (*16*).

The second hypothesis for the origin of the pseudogap involves preformed pairs(*3, 17–19*). In this scenario, electron pairs, or fluctuations thereof, exist above $T_c$ and cause a depletion in the density of states with energy $\Delta_{PG}$. The pseudogap phase is thus a precursor of the superconducting state, but with the paired electrons lacking phase coherence above $T_c$. This scenario is often called the one-gap scenario, and within it, the meaning of the kink-like signature is unclear. Support for this scenario stems from many tantalizing but controversial signatures of preformed pairs across the phase diagram at temperatures up to $T_{PG}$(*1, 5, 17*), while others are observed at temperatures between $T_c$ and $T_{PG}$ (*20*).

From the above, it is clear that measuring the energy at which Cooper pairs emerge can unambiguously reveal which of the two physical scenarios describes the pseudogap phase.

In this article, based on shot-noise measurements, we unequivocally show that the local-order hypothesis can be excluded. Instead, the pseudogap energy $\Delta_{PG}$ is clearly associated with pairing, up to energies of more than 70 meV. It follows that condensation is the barrier to achieving higher $T_c$ in cuprate superconductors, that pairing is heterogenous, and that the pseudogap cannot solely stem from local order.

Shot noise is a direct method to detect electron pairing (*21–23*). In a tunnel junction, the tunneling process follows Poissonian statistics, owing to the discrete nature of the charge carriers. This leads to current fluctuations known as shot noise. At low temperatures and transparencies, the shot noise power is $S_I = 2q|I|$, where $I$ is the tunneling current and $q$ is the effective charge transported (*24*). When tunneling into a superconductor, Andreev reflections contribute to the current. These effectively transport $2e$, so that the noise from this process is twice the noise from single quasiparticle tunneling, which is the tell-tale signature of pairing. We recently combined shot noise spectroscopy with scanning tunneling microscopy (STM) to achieve noise-STM with high sensitivity (*22, 25*). Noise-STM allows local measurements, provides a clean vacuum barrier that does not suffer from the chemistry of solid-state barriers, and enables a simpler interpretation of shot noise than mesoscopic tunnel junctions, which are prone to artifacts (*26*, *27*) (Supplementary Information, section 8). It further allows us to simultaneously measure the local density of states (LDOS) at a sample surface through its spectroscopic mode. We have successfully used this technique to detect electron pairing in s-wave superconductors (*22*, *28*); here, we bring it, for the first time, to the much more challenging nodal superconductors.

The nodal character of cuprate high-temperature superconductors implies that the total tunneling current at energies below the maximum gap is carried by both quasiparticles and Andreev reflections. The former contributes strongly to the total noise because quasiparticle tunneling has a higher probability at low transparency than Andreev tunneling (*26*). This decreases the expected effective charge equivalent in noise experiments to a value between 1e and 2e. In Fig. 1C, we show simulations of the expected noise for both s-wave and d-wave superconductors using the framework of the Blonder-Tinkham-Klapwijk (BTK) model, see Supplementary Materials for details. The simulations are done for transparencies around $\tau \sim 10^{-3}$. Indeed, the effective charge expected from a d-wave superconductor is much lower than the 2e value typical for a s-wave superconductor (Fig. 1C), and significant upgrades on our technique were necessary to achieve the needed sensitivity.

We perform simulations (Fig. 1F) for both of the hypotheses about the origin of the pseudogap that are put forward at the beginning of the paper: (1) a superconducting gap with energy $\Delta_{\text{kink}}$, inside a larger gap induced by local order (Fig. 1D) and (2) the case where the pseudogap is associated with pairing (Fig. 1E). Our simulations do not yield robust quantitative numbers (because small changes in the number of nodal quasiparticles can significantly change the value of the effective charge measured by shot noise). Nonetheless, they show robust qualitative features: a step in the effective charge at the onset energy of pairing, which we therefore call $E_{\text{pair}}$ (dashed vertical lines in Fig. 1F), followed by a continuous increase of the effective charge towards lower bias. Thus, the question of which microscopic scenario is behind the pseudogap is transformed to: Does a step in the effective charge occur at the pseudogap energy or at the kink energy?

We perform the experiments in our homemade, ultra-stable STM at $T = 4.2$ K and in modified commercial STMs at 2.3 K and 0.3 K (see Supplementary Materials) with samples of several different doping concentrations. We first measure shot noise at $T = 4.2$ K on an underdoped Bi-2212 sample with $T_c = 60$ K (UD60K), which has clearly separated energy scales $\Delta_{\text{kink}}$ and $\Delta_{\text{PG}}$ (Fig. 2A). Fig. 2 shows the key result of our study: the effective charge $q$ as a function of bias voltage, compared with the differential conductance $dI/dV$ measured at the same lateral position. We find that $q = 1e$ at high bias voltages, i.e., for $|V| > \Delta_{PG}/e$. However, at a bias voltage of $|V| \sim \Delta_{PG}/e$, the noise increases, leading to a small step in the effective charge at the energy at which the outer peaks – corresponding to the pseudogap – in the differential conductance spectrum are observed. With decreasing bias voltage $|V| < \Delta_{PG}/e$ the effective charge increases further, which is consistent with our simulations for $\Delta_{\text{PG}} = E_{\text{pair}}$. Hence, pairing is present up to the pseudogap energy. In other words, the comparison of Fig. 2A and B, reveals that pairing starts at the pseudogap energy and therefore, that the pseudogap is related to pairing. Note that we define $\Delta_{PG}$ as the energy that shows a maximum in the differential conductance spectra.

Our result excludes the possibility of a pure two-gap scenario, i.e., the pseudogap being caused solely by local order. However, we emphasize that our results do not rule out the possibility of a pseudogap that is related to both, pairing and local order. It has often been conjectured that the many orders in the cuprate are intertwined (*11*, *29*, *30*), so that pairing and other orders form one gap. Of particular interest might be the recently measured pair density wave (*31*, *32*), as its symmetry allows naturally to relate pairing with other local orders.

To explore spatial heterogeneity, we next measure noise at additional positions with different $\Delta_{PG}$ (Fig. 3). At each location, we measure the pseudogap energy $\Delta_{PG}$ and the step energy $E_{pair}$ (gray lines in Fig. 3). When plotting them against each other, a clear correlation becomes apparent (Fig. 3C). The slope is 1, confirming that the pairing energy is related to the pseudogap energy. We confirm this with data from samples with different $T_c$: underdoped samples with $T_c = 70$ K (UD70K) and optimal doped samples with $T_c = 91$ K (OP91K); for details, see Fig. S9 and S10. This brings us to our second conclusion: our data shows that the pairing energy in cuprate high-Tc superconductors is large compared to $T_c$ and strongly heterogenous.

We further want to address the possibility that the noise originates from other sources. There could be random-telegraph noise stemming from changes in the atomic configuration (*33–36*) or $1/f$ noise from tunneling processes(*37*). It is unlikely that these sources contribute to the measured noise, as the shape of the noise spectra have a particular shape that agrees with theoretical simulations and our amplifier is designed and calibrated to reduce other noises (see Supplementary Materials). Still, to fully exclude these sources, we perform noise measurements under an external magnetic field on OP91K and a sample with $T_c = 58$ K (UD58K). Noise from other sources should be independent of the field, while shot noise from pairing should decrease significantly because the gap structure allows quasiparticles to exist far from vortices, similar to the measurements performed on NbSe$_2$(*38*). We measured the noise spectra at the same locations as before under an external magnetic field $B$=0 T and $B$=1 T, and at various additional locations at $B = 1.4$ T (UD58K) and $B = 6$ T (OP91K). We indeed observed that the noise approaches $1e$ at higher fields (Fig. 4), further supporting our conclusions.

Our work has consequences for the understanding of pairing and coherence. The pseudogap energy scale increases with decreasing doping concentration (see Fig. 1A). However, as the hole doping concentration further decreases, $T_c$ reaches a maximum value after which it decreases and eventually vanishes. Clearly, superconductivity cannot be limited by pairing; it must come from a lack of phase coherence between the Cooper pairs. An intuitive and well-known model describing the loss of long-range coherence consists of superconducting islands separated by Josephson junctions; at higher temperatures, these junctions cannot sustain coherence between islands, making the material resistive although short-range pairing remains. This has been observed in patterned conventional and unconventional superconductors (*39*, *40*) and in overdoped cuprates, which form intrinsic puddles (*41*). However, we do not observe clear puddles in local density of states mappings. It is tempting to associate the kink energy with coherence. Theoretical calculations of how coherence could be reflected in the differential conductance spectra would be highly desirable.

A remaining question is whether the increased noise up to $\Delta_{PG}$ appears at temperatures above $T_c$. This temperature range is currently not accessible with our technique as there are too many quasiparticles, and the sensitivity of our amplifier decreases with increasing temperature. However, we emphasize that our work already conclusively demonstrates a connection between the pseudogap and pairing. We use the large gap $\Delta_{PG}$ as observed in the differential conductance measurements as the defining feature. However, it is important to consider that this spectral gap might not be exactly the same as the one measured by other techniques, i.e., photoemission spectroscopy has observed distinct gaps in different momentum space directions (*2*). Naming features such as $\Delta_{PG}$ in a consistent fashion across different techniques is challenging as definitions vary and the crossovers are broad. However, our data shows pairing energies higher than 70meV for certain spatial locations, which exceeds any superconductivity-related spectroscopic gap measured in this material, indicating that the pairing onset is given by the pseudogap.

In conclusion, our data shows that Cooper pairs in cuprates are present up to the pseudogap energy, which is significantly larger than the expected superconducting gap energy. The expectation from BCS is that this should result in higher $T_c$. A natural implication is that superconductivity in cuprates is limited by phase coherence, not pairing. We also observe that the pairing energy is spatially heterogeneous. Our findings may inspire hope for materials with higher transition temperatures and provide a possible path forward for future research.

**Acknowledgements:** We acknowledge Steven Kivelson, Zhi-Xun Shen, and Yu He for discussions. This work was supported by the European Research Council (ERC StG SpinMelt and ERC CoG PairNoise). I. S. and M. O. L. were supported by the European Research Council (Horizon 2020 "FRACTAL", 865570). Y. M. B. was supported by European Space Agency (ESA) under ESA CTP Contract No. 4000130346/20/NL/BW/os. The work at BNL was supported by the US Department of Energy, office of Basic Energy Sciences, contract no. DOE-sc0012704.

**Data and materials availability:**

All data used to generate the figures in the main text and the supplementary information is available upon request to the authors.

**Code availability:**

The code used for this project is available upon request to the authors.

# Figures

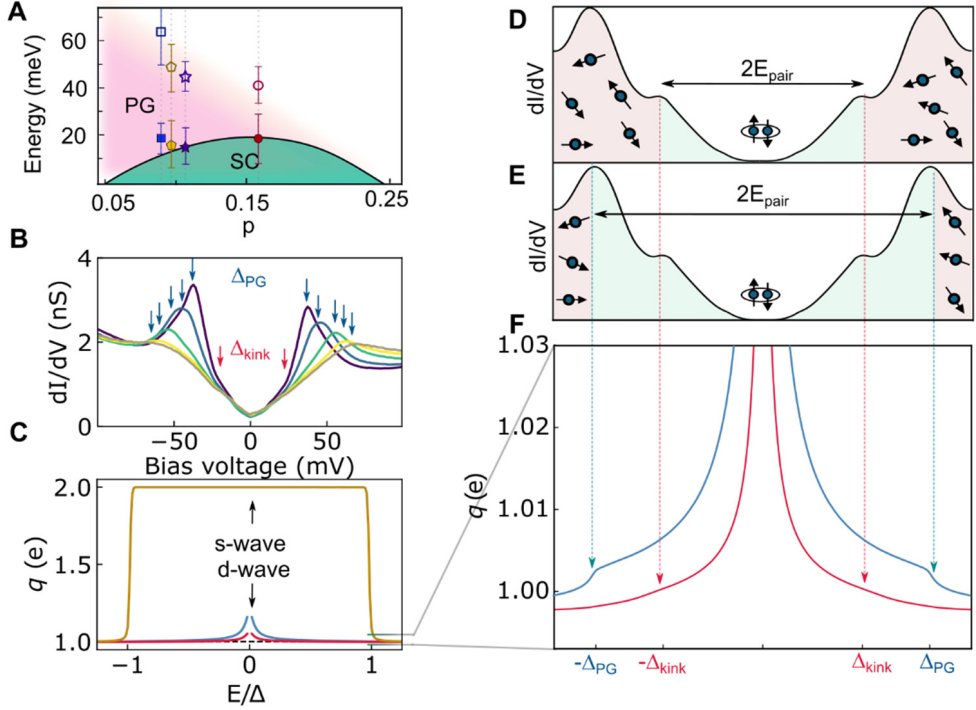

**Fig. 1. Spectroscopic signatures of pseudogap and superconductivity in high-temperature superconductors.** (**A**) Phase diagram of energy as function of hole doping for cuprate superconductors, based on reference(*17*). The colored markers indicate $\Delta_{PG}$ and $\Delta_{kink}$ for $Bi_2Sr_2CaCu_2O_{8+\delta}$ samples with different $T_c$ that we measured: OP91K (red circles), UD70K (purple stars), UD60K (brown pentagons) and UD58K (blue squares). Samples are named by their $T_c$ (in kelvin) with the prefix UD for underdoped and OP for optimally doped. (**B**) Typical differential conductance spectra of a UD60K sample measured at $T = 4$ K. $\Delta_{PG}$ and $\Delta_{kink}$ are indicated by arrows. (**C**) Effective charge *q* vs *E/Δ* calculated for s-wave (yellow line) and d-wave (red and blue lines) superconductors. (**D-F**) *Differential conductance sketches* and *q* vs *bias voltage* simulations for two possible physical origins of the pseudogap: $\Delta_{PG}$ is related to local orders and $\Delta_{kink}$ is related to pairing (panel (**D**) and red curve in panel (**F**)), and $\Delta_{PG}$ is related to pairing (panel (**E**) and blue curve in panel (**F**)). $\Delta_{PG}$ and $\Delta_{kink}$ are indicated by blue and red dashed lines, respectively, while $2E_{pair}$ is indicated by black arrows in both scenarios. The parameters used for the effective charge simulations are indicated in the Supplementary Materials .

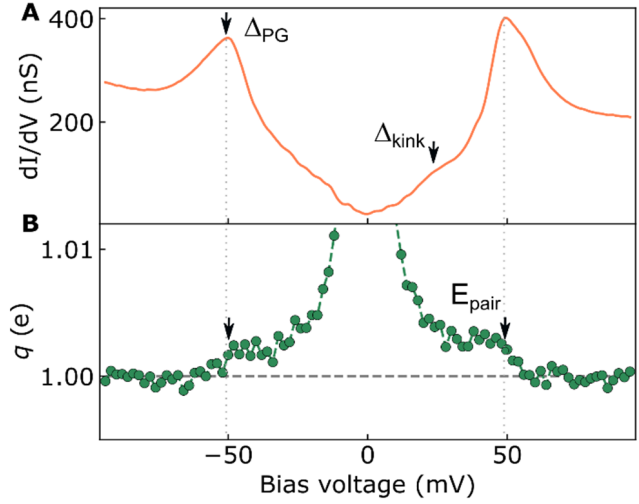

**Fig. 2. Pairing up to the pseudogap energy.** (**A**) Typical differential conductance spectrum of an underdoped $Bi_2Sr_2CaCu_2O_{8+\delta}$ ($T_c = 60$ K) with set-up conditions $V = 94$ mV and $I = 18.8$ nA. $\Delta_{PG}$ is determined by the peak position, indicated by arrow and dashed lines. (**B**) Effective charge $q$ at this position measured by shot noise with a constant tunneling resistance of $R_j = 5$ MΩ at both positive and negative biases (green dots). $q$ clearly increases with decreasing bias voltage $|V|$, and shows a small step structure which is consistent with our simulations. The pairing energy $E_{pair}$ is indicated by arrows. The process for extracting $E_{pair}$ is explained in the Supplementary Materials.

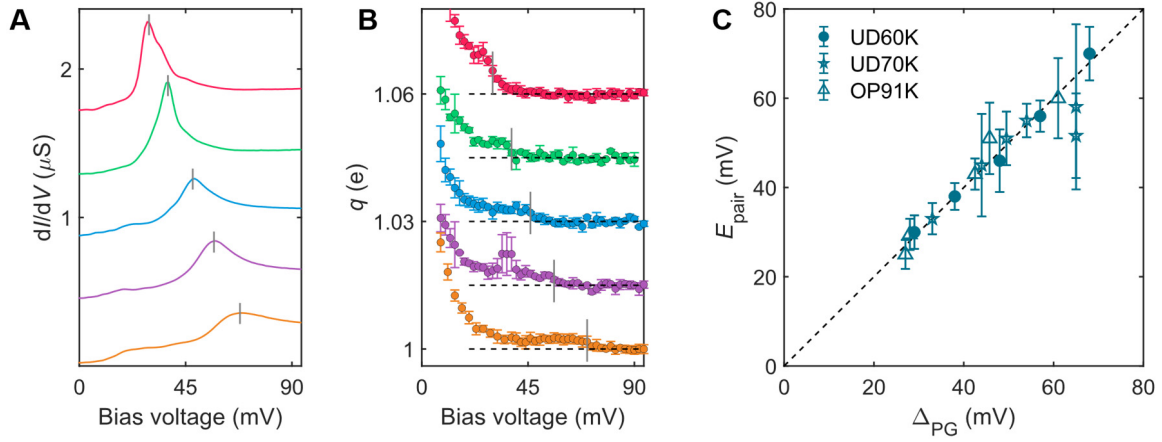

**Fig. 3. *S*patial heterogeneity of the pairing energy.** (**A**) *dI/dV* spectra at different positions on the UD60K sample with set-up conditions $V = 94$ mV and $I = 18.8$ nA. The results are shifted by a vertical offset. Vertical gray lines indicate the values of $\Delta_{PG}$. (**B**) Effective charge *q* as measured by noise-STM at the same positions as in (**A**), with a constant tunneling resistance of $R_j = 5$ MΩ. Each point is measured and averaged multiple times, with the standard deviation plotted as an error bar. Results are shifted with a constant offset and black dashed lines indicate $q = 1e$. Vertical gray lines indicate the value of $E_{pair}$. For details on the extraction of $E_{pair}$, see Supplementary Materials. (**C**) Summary of $E_{pair}$ vs. $\Delta_{PG}$ for different samples. The black dashed line represents $E_{pair} = \Delta_{PG}$, and the results clearly follow this behavior. Detailed information about the uncertainty of the results can be found in the Supplementary Materials and Fig. S3.

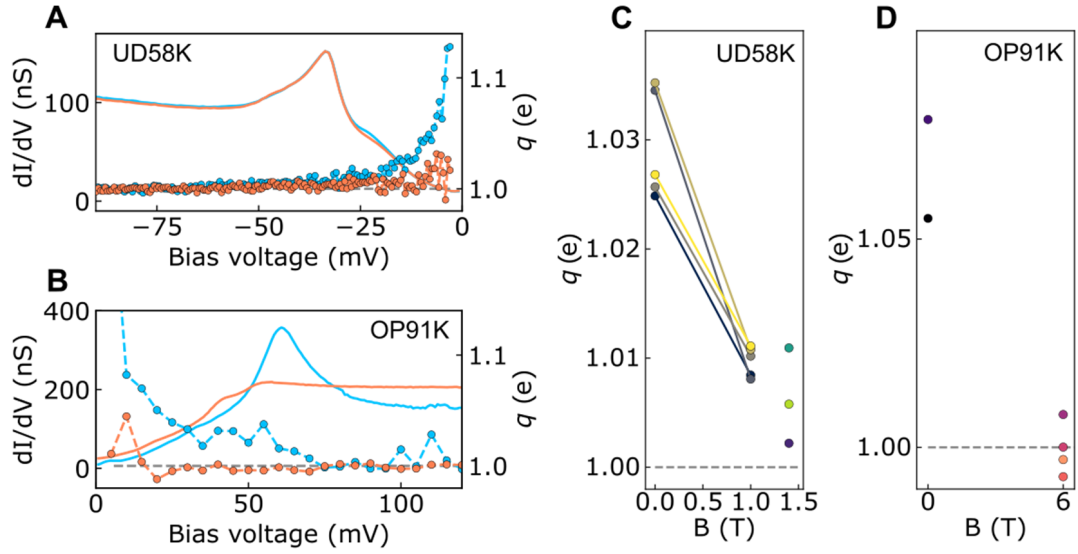

**Fig. 4. Magnetic field dependence.** (**A**) $dI/dV$ spectra (solid lines) and effective charge $q$ (filled circles) as a function of bias voltage for $B$ = 0 T (blue) and 1 T (red) on a single position of the UD58K sample. (**B**) Similar results on OP91K sample for 0 T (blue) and 6 T (red), for unequal positions. (**C, D**) Average $q$ versus magnetic field for different positions on UD58K and OP91K, respectively. The values of $q$ are averaged for a range of bias voltages from $\Delta_{PG}/6$ to $\Delta_{PG}/2$. Results obtained from the same positions are connected by solid lines. Dashed lines indicate $q = 1e$. Detailed results of $q$ versus voltage for all the positions are plotted in the Fig. S11-S13.

Supplementary Materials for

# Equivalence of pseudogap and pairing energy in a cuprate high-temperature superconductor

Jiasen Niu†, Maialen Ortego Larrazabal†, Thomas Gozlinski, Yudai Sato, Koen M. Bastiaans, Tjerk Benschop, Jian-Feng Ge, Yaroslav M. Blanter, Genda Gu, Ingmar Swart, Milan P. Allan*.

† These authors contributed equally. * Corresponding author: milan.allan@lmu.de

**Materials and methods:**

1. Calculation of shot noise in N-I-S junctions using the BTK formulas

2. Analytical calculations of shot noise in N-I-S junctions (s-wave and d-wave)

3. Shot noise extraction and calibration

4. Excluding the noise from other sources

5. Sample preparation and STM measurements

6. Extracting $E_{pair}$ from shot-noise measurements

7. Extracting $\Delta_{PG}$ and $\Delta_{kink}$ from the differential conductance measurements

8. Why shot noise in mesoscopic junctions does not reflect pairing.

**Figures:**

Fig. S1. BTK simulation of a d-wave superconductor for different values of $\alpha$

Fig. S2. Effective charge measurement on sample UD60K and Au(111).

Fig. S3. Uncertainty of $q$

Fig. S4. Topographic image of OP91K sample

Fig. S5. Extracting $E_{pair}$ from simulations

Fig. S6. Extracting $E_{pair}$ and uncertainty of $E_{pair}$ for UD60K

Fig. S7. Absolute value of $q$ varies with amplifier calibration range

Fig. S8. Extracting $\Delta_{PG}$ and $\Delta_{kink}$ from tunneling spectra map

Fig. S9. More results of OP91K at $B = 0$ T

# Materials and Methods

## 1. Calculations of shot noise in N-I-S junctions using the Blonder-Tinkham-Klapwijk formulas

In the main text, we show the calculation results of the effective charge, which allows us to estimate the expected charge enhancement from our measurements (Fig.1). These numerical calculations are performed using a phenomenological approach, in which the relevant gap parameters are obtained from fitting the measured dI/dV spectra and then used in a scattering matrix approach to calculate the effective charge. In this section, we first introduce the Blonder-Tinkham-Klapwijk (BTK) formulas(1–3) that describe the scattering problem and then generalize them to model a d-wave gap(4).

**The BTK model.** The BTK model(1–3) is based on the Bogoliubov–de Gennes (BdG) equations and describes the tunneling current for a normal metal–superconductor (N-S) junction with an arbitrary barrier strength in the framework of scattering matrix formalism. The superconductor is assumed to be of s-wave pairing symmetry. There are two possible tunneling processes at an N-S interface: The Andreev reflection process and the normal tunneling process. The Andreev reflection coefficient $A$ and the normal state reflection coefficient $B$ take the following form:

|  | $A$ | $B$ |
|---|---|---|
| $\varepsilon < \Delta$ | $\left\|\dfrac{u_0 v_0}{\gamma}\right\|^2$ | $1 - A$ |
| $\varepsilon > \Delta$ | $\left\|\dfrac{u_0 v_0}{\gamma}\right\|^2$ | $\left\|-\dfrac{u_0^2 - v_0^2}{\gamma}(Z^2 + iZ)\right\|^2$ |

with the coefficient $\gamma = u_0^2 + Z^2(u_0^2 - v_0^2)$, the BCS factors $u_0^2 = 1 - v_0^2 = \frac{1}{2}\{1 + [(\varepsilon^2 - \Delta^2)/\varepsilon^2]^{1/2}\}$, and the superconducting gap $\Delta$. The parameter $\varepsilon$ includes quasiparticle lifetime broadening $\Gamma$ and the energy $E$ as $\varepsilon = E - i\Gamma$. The dimensionless barrier strength $Z$ describes the height of the barrier and is related to the transmission $\tau$ through the barrier in the normal state as $\tau = 1/(1 + Z^2)$.

**Current and noise power for a s-wave gap.** The current $I^s$ and shot noise power $S^s$ for an s-wave superconductor are then calculated as(1, 5)

$$I^s = 2N_0 ev\mathcal{A} \int_{-\infty}^{\infty} [1 + A - B][f(E - eV) - f(E)]dE \qquad (1)$$

$$S^s = 4N_0 e^2 v\mathcal{A} \int_{-\infty}^{\infty} [B(1 - B) + A(1 - A) + 2AB][f(E - eV) - f(E)]dE, \qquad (2)$$

where $N_0$ is the normal-state density of states at Fermi energy, $e$ is the electron charge, $v$ is the Fermi velocity, $\mathcal{A}$ is the effective neck cross-sectional area, $V$ is the applied bias voltage, $h$ is the Planck constant and $f(E) = \frac{1}{1+ e^{E/k_B T}}$ is the Fermi-Dirac distribution with the Boltzmann constant $k_B$ and temperature $T$. The effective charge $q$ is extracted from the ratio between these two quantities, $q = S/2eI$. Note that we do not need to determine the proportionality constant $N_0 e^2 v \mathcal{A}$ as it cancels out in the ratio.

**Model for tunneling into d-wave superconductors.** Next, we generalize the BTK formulas for an arbitrary d-wave gaps [4]. In a later section, we treat the problem analytically, here we show how to numerically obtain estimates for different scenarios. We model the system with a d-wave gap as a superposition of s-wave gaps at different points in momentum space along the Fermi surface (line in 2D). Each s-wave gap has magnitude $\Delta_\mathbf{k}$, where $\Delta_\mathbf{k} = \frac{\Delta_0}{2}\left(\cos(\pi k_{x0}) - \cos(\pi k_{y0})\right)$, where each $\mathbf{k} = (k_{x0}, k_{y0})$ is a vector on the Fermi surface in the first Brillouin zone, and $\Delta_0$ is the maximum superconducting gap.

We then use the BTK model to calculate the current contributions for each s-wave gap, $I_\mathbf{k}^s$ at momentum k on the Fermi surface. However, we need to weight the s-wave gaps correctly to adjust for density of states effects and the length of the Fermi-surface line. We first introduce the k-dependent normal state density of states for each $\mathbf{k}$-value of the Fermi surface $N_{0\mathbf{k}} = \frac{1}{|\nabla_\mathbf{k} E(\mathbf{k})|}$ following Ref. (6). We then calculate the total current as

$$I^d = \frac{1}{N_0} \oint N_{0k} I_\mathbf{k}^s ds, \quad (3)$$

where the integral runs over the Fermi surface (which is a curve in our case), and $ds$ is the infinitesimal Fermi surface (line) element. We use the same procedure to calculate the noise $S^d$.

**Numerical implementation.** Fig.1F shows simulations for two different scenarios: local orders scenario (red curve) and pairing scenario(blue curve). For the pairing scenario presented in Fig. 1E (blue curve in Fig. 1F), where the gap is caused only by a BCS-like pairing gap, we use a simple d-wave gap with parameters roughly corresponding to the pseudogap in our sample ($\Delta_0^{PG} = 60$ meV, $\Gamma = 0.01 * \Delta_0^{PG}, \tau = 2.5 \times 10^{-3}, T = 4.2\ K$). We use the Fermi surface and the k-resolved normal state density of states from a tight-binding model described in Ref. (7). We then calculate the noise and current by numerical integration over 900 points on the Fermi surface.

In the local orders scenario (Fig.1D and red curve in Fig. 1F), the pseudogap with amplitude $\Delta_0^{PG}$ is not caused by pairing and hence does not contribute to Andreev reflections, while a smaller superconducting gap with amplitude $\Delta_0^{SC}$ (chosen as two thirds of $\Delta_0^{PG}$, roughly the "kink" energy) is associated to pairing. We separate the Fermi surface into two parts. One part is around the nodal point, where $\Delta_\mathbf{k} < \Delta_0^{SC}$. This is related to the smaller superconducting gap. The other part is near the anti-nodal point, where $\Delta_0^{PG} > \Delta_\mathbf{k} > \Delta_0^{SC}$. Thus, only the k-values that fulfil the condition $\Delta_\mathbf{k} < \Delta_0^{SC}$ are related to pairing and yield Andreev reflection.

For $\Delta_{\mathbf{k}} > \Delta_0^{SC}$, there is no Cooper pair contribution to the effective charge, leading to $q = 1 - \tau$.

Extended Data Fig. 1 shows the effect of the quasiparticle lifetime broadening in the differential conductance and effective charge.

## 2. Analytical calculations of shot noise in N-I-S junctions

In order to better understand the results from the BTK simulations, we look at the expected effective charge from N-I-S junctions analytically at zero temperature, $T = 0$ K. These calculations are done for both s-wave and d-wave cases and agree with our numerical BTK calculations in the parameter space where the analytical calculations are performed.

**Normal metal – s-wave superconducting interface.** As a prerequisite to the calculation, we need an expression for shot noise at an interface between a normal metal and an s-wave superconductor. This expression can be easily found in the literature for $eV \ll \Delta$(*8*). However, here we will need the shot noise also for voltages of the order of the gap or bigger, and this section provides such expressions. For simplicity, we consider only one transport channel.

The starting point are the expressions through the interface which were derived by Blonder, Tinkham, and Klapwijk(*1*). If we count energies $E$ from the Fermi energy, for $0 < E < \Delta$, an electron can not be transmitted to the superconductor, and it can only be reflected as an electron (normal reflection) or as a hole (Andreev reflection). Normal reflection does not lead to charge transfer into the superconductor, whereas Andreev reflection is accompanied with the charge transfer $2e$. The probability of Andreev reflection is(*1, 8*)

$$R_A(E) = A(E) = \frac{\tau^2}{(1+R)^2 - 4RE^2/\Delta^2}, 0 < E < \Delta, \qquad (4)$$

where $\tau$ and $R = 1 - \tau$ are the transmission and reflection coefficients of the barrier, respectively, if the whole system is in the normal state. In the following, we assume that $\tau$ and $R$ do not depend on the energy $E$. This is a common assumption in the field and makes sense if the height of the energy barrier is much larger than the superconducting gap.

For $E > \Delta$, there are two additional processes in which an electron can be transmitted as an electron-like and a hole-like quasiparticle into the superconductor. Out of these four processes, Andreev reflection is still accompanied by the charge transfer of $2e$, whereas normal transmission results in a charge transfer of $1e$. The other two processes do not result in charge transfer. The probability of Andreev reflection is(*1*)

$$R_A(E) = \frac{\tau^2 \Delta^2}{\left[\tau E + (1+R)\sqrt{E^2 - \Delta^2}\right]^2}, E > \Delta, \qquad (5)$$

and the probability of normal transmission reads(*1*)

$$\tau_N(E) = \frac{2\tau\sqrt{E^2 - \Delta^2}\left(E + \sqrt{E^2 - \Delta^2}\right)}{\left[\tau E + (1+R)\sqrt{E^2 - \Delta^2}\right]^2}, E > \Delta. \tag{6}$$

Eq. (4)-(5) are the same as the expressions in Table S1. Using these probabilities, we can calculate the electric current with Eq. (4). For $eV < \Delta$, all current proceeds via the Andreev reflection,

$$I_A = \frac{2G_Q}{e} \int_0^{eV} R_A(E) dE = \frac{G_Q \tau^2 \Delta}{2e\sqrt{R}(1+R)} \ln \frac{(1+R)\Delta + 2\sqrt{R}eV}{(1+R)\Delta - 2\sqrt{R}eV}, \tag{7}$$

where $G_Q = e^2/(\pi\hbar)$ is the conductance quantum which includes the factor of 2 due to the spin. Note that in the limit $eV \ll \Delta$ and $\tau \ll 1$, Eq. 7 reduces to the familiar expression(9) $I_A = \frac{1}{2} G_Q \tau^2 V$.

For $eV > \Delta$, we have three contributions to the current. For energies below $\Delta$, the contribution is from Andreev reflections, and it equals to Eq. 7 evaluated at $eV = \Delta$. For energies above $\Delta$, the current originates from both Andreev reflections and tunneling of the quasiparticles and can be evaluated by using the Landauer formula(9). We thus have

$$I = I_A(eV = \Delta) + \frac{2G_Q}{e} \int_\Delta^{eV} R_A(E) dE + \frac{G_Q}{e} \int_\Delta^{eV} \tau_N(E) dE. \tag{8}$$

These integrals can not be evaluated analytically, but we can estimate them in the limit of interest $\tau \ll 1$. In that case, the second integral approximately equals $I_A(eV = \Delta)$ for voltages slightly away from $\Delta$, $eV - \Delta \gtrsim \Delta \tau^2$. The normal component of the current for $\tau \ll 1$ is

$$I_N \equiv \frac{G_Q}{e} \int_\Delta^{eV} \tau_N(E) dE = \frac{G_Q \tau}{2e} \left[eV - \Delta + \sqrt{(eV)^2 - \Delta^2}\right] \tag{9}$$

Similarly, for the current noise density at zero frequency and $eV < \Delta$, we have

$$S_A = 8G_Q \int_0^{eV} R_A(E)[1 - R_A(E)] dE$$

$$= 2G_Q \tau^2 \Delta \left\{ \begin{array}{l} \frac{1}{\sqrt{R}(1+R)} \left[1 - \frac{\tau^2}{2(1+R)^2}\right] \\ \times \ln \frac{(1+R)\Delta + 2\sqrt{R}eV}{(1+R)\Delta - 2\sqrt{R}eV} - \frac{2\tau^2}{(1+R)^2} \frac{eV\Delta}{(1+R)^2\Delta^2 - 4R(eV)^2} \end{array} \right\}. \tag{10}$$

For $eV \ll \Delta, \tau \ll 1$ we recover the familiar expression(8, 9) $S_A = 2G_Q \tau^2 eV$. Defining the Fano factor as $F \equiv S/2eI$, we get $F = 2$ in this limit, corresponding to the notion that Andreev reflection is accompanied by a charge transfer of $2e$. Generally, assuming $\tau \ll 1$ but not $eV \ll \Delta$, we see from Eq. (7) and Eq. (10) that the Fano factor remains close to 2 for the whole range of voltages. In particular, for $eV = \Delta$, it equals to $F = 2\left[1 - \frac{1}{2(2\ln 2 - \ln \tau)}\right]$.

For the energies above the gap we have, similarly to Eq.8

$$S = S_A(eV = \Delta) + 8G_Q \int_\Delta^{eV} R_A(E)[1 - R_A(E)]dE$$
$$+ \frac{G_Q}{e} \int_\Delta^{eV} \tau_N(E)[1 - \tau_N(E)]dE. \quad (11)$$

For $\tau \ll 1$, the second term is approximately equal to the first one as soon as we are slightly above $\Delta$, $eV - \Delta \gtrsim \Delta \tau^2$, and the last integral is $S_N = 2eI_N$, since $\tau_N \ll 1$ for all energies. We thus have $S \approx 2S_A(eV = \Delta) + 2eI_N$.

Note that the Fano factor is 1 for $eV \gg \Delta$, when the charge transfer is dominated by normal reflections. We thus have a crossover from $F = 2$ to $F = 1$ with increasing voltage. The width of the crossover can be estimated from the condition $I_N \sim I_A(eV = \Delta)$, and we see that the Fano factor drops from 2 to 1 very sharply, between $eV = \Delta$ and approximately $eV = \Delta[1 + \left(\frac{\tau^2}{2}\right) ln^2(2/\tau)]$ with $\tau \ll 1$.

**Normal metal – d-wave superconducting interface.** We now use the formalism above to describe current and noise at the interface between a normal metal and a d-wave superconductor. In a d-wave superconductor, the gap depends on the direction $\theta$ in the $xy$-plane (perpendicular to the tunneling direction) such that it stays invariant under $\frac{\pi}{2}$ rotations and vanishes in certain directions. We approximate it by $\Delta(\theta) = \Delta_0 \cos 2\theta$. Whereas this is not an exact expression for the order parameter in cuprate superconductors(7) used above for the numerical simulations, it has a correct symmetry and will allow us to gain an analytical understanding of the Fano factor.

To proceed, we assume that the incident electrons have all possible transverse components of momentum and thus are characterized by a uniform distribution of angles $\theta$. The measured current and noise are then obtained as averages,

$$\binom{I_d}{S_d} = \frac{4}{\pi} \int_0^{\frac{\pi}{4}} d\theta \binom{I[\Delta(\theta)]}{S[\Delta(\theta)]}, \quad (12)$$

where I and S are the current and noise for the s-wave superconducting interface found in the previous section. Note that the gap vanishes for $\theta = \pi/4$, and thus for any voltage V we have directions when only Andreev reflection is possible ($0 < \theta < \theta_0$) and directions when both Andreev reflection and normal transmission are possible ($\theta_0 < \theta < \pi/4$). Here, $2\theta_0 = \frac{\pi}{2} - \arcsin eV/\Delta_0$.

In the following, we consider the limit of low voltages, $eV \ll \Delta_0$. Then $\theta_0 = \pi/4 - eV/(2\Delta_0)$.

We first note that the calculation shows that in the limit $\tau \ll 1$, $eV \ll \Delta_0$ the main contribution of Andreev reflection for both current and noise is provided by the angles $0 < \theta < \theta_0$, $i.e.$ the contribution of Andreev reflection processes above the gap is not significant.

The Andreev contribution for the current and noise by $0 < \theta < \theta_0$ in the limit of interest $\tau \ll 1$, $eV \ll \Delta_0$ reads,

$$I_{Ad} \equiv \frac{4}{\pi} \int_0^{\theta_0} I_A(\theta) d\theta = \frac{1}{2} G_Q \tau^2 V ; \tag{13}$$

$$S_{Ad} \equiv 4/\pi \int_0^{\theta_0} S_A(\theta) d\theta = 2 G_Q \tau^2 eV . \tag{14}$$

In the same limit, the normal contribution to the transport becomes

$$I_{Nd} \equiv \frac{4}{\pi} \int_0^{\theta_0} I_N(\theta) d\theta = \frac{1}{2\Delta_0} G_Q \tau eV^2 \left(\frac{1}{2} + \frac{1}{\pi}\right) ; \tag{15}$$

$$S_{Nd} \equiv \frac{4}{\pi} \int_0^{\theta_0} S_N(\theta) d\theta = \frac{1}{\Delta_0} G_Q \tau (eV)^2 \left(\frac{1}{2} + \frac{1}{\pi}\right). \tag{16}$$

Calculating the Fano factor, we obtain

$$F = \frac{S_{Ad} + S_{Nd}}{2e(I_{Ad} + I_{Nd})} = \frac{2\tau\Delta_0 + eV\left(\frac{1}{2} + \frac{1}{\pi}\right)}{\tau\Delta_0 + eV\left(\frac{1}{2} + \frac{1}{\pi}\right)}. \tag{17}$$

This expression describes a smooth crossover from the value $F = 2$ at $V = 0$ to $F = 1$ for $eV \gg \tau\Delta_0$. The typical voltages at which the crossover occurs are $eV \sim \tau\Delta_0$ (note that the numerical factor of the order one in this relation is dependent on model assumptions such as the angular dependence of the order parameter). This smooth voltage dependence of the Fano factor for the normal metal - d-wave superconductor interface is in contrast with the sharp voltage dependence of the Fano factor for the normal metal - s-wave superconductor interface and is related to the fact that for the d-wave case at every voltage we have both normal transmission (responsible for the $e$ charge transfer) and Andreev reflection (responsible for the $2e$ charge transfer). This conclusion is qualitatively consistent with our BTK simulations and with our measurements at small bias ($eV \ll \Delta_0$). The analytical simulation is calculated in the limit where $eV \ll \Delta_0$. Therefore, it does not work for the energy regime close to $\Delta_0$, where our measurements and BTK simulations indicate a step-like behavior.

## 3. Shot noise extraction and calibration

The shot noise is measured by a homemade cryogenic MHz amplifier system including a LC resonance circuit, a high electron mobility transistor (HEMT)-based high frequency amplifier, a room temperature commercial amplifier (FEMTO HAS-X-1-40) and a spectrum analyzer (Zurich MFLI or HF2LI). Details of the low temperature amplifier can be found in the previous work(*10*). Throughout the experiments, we measure the noise at different bias voltage $V$, with constant junction resistance $R_j$, which is controlled by the STM feedback system. For each bias, we will obtain a current $I = \frac{V}{R_j}$ and shot noise power $S_I^{shot} = 2qI *$ coth $\left(\frac{qV}{2k_BT}\right)$, where $q$ is the effective charge, $T$ is the temperature and $k_B$ is the Boltzmann constant. The shot noise will be first translated from current noise $S_I^{shot}$ to voltage noise $S_V^{shot}$ by the LC resonance circuit and then amplified by low temperature and room temperature amplifiers. Thus, the noise measured by spectrum analyzer, $S_V^m$, is

$$S_V^m = G^2 * |Z_{tot}^2| * (S_I^{shot} + BG_1^I) + BG_0^V, \qquad (18)$$

where $G$ is the total gain of the amplifier chain, $BG_0^V$ is the effective voltage noise background of the amplifier chain and the spectrum analyzer, $Z_{tot} = Z_{LC}//R_j^{diff}$ is the total impedance of LC resonator $Z_{LC}$ and STM junction differential resistance $R_j^{diff}$, and $BG_1^I$ is the effective current noise background of the low-temperature amplifier and the resonator circuit. Then, the effective charge can be calculated by

$$2qI * \coth\left(\frac{qV}{2k_BT}\right) = S_I^{shot} = (S_V^m - BG_0^V)/(G^2|Z_{tot}^2|) - BG_1^I. \qquad (19)$$

To extract an accurate value of the effective charge, our amplifier system is carefully calibrated. $G$ and $BG_1^I$ are first calibrated on a Au(111) surface and results are shown in the Extended Data Fig. 2. After this, for each measurement position on BSCCO, we will use data at bias larger than the gap to calibrate the $G$ and $BG_1^I$ again to avoid the tiny change (< 1%) due to the fluctuations in the environment. In addition, $Z_{tot}$ is calibrated for data at each bias using the shape of the resonance peak in the frequency domain. Details can be found in a previous work(*11*). This calibration removes the artifact from the fluctuations in the environment capacitance. $BG_0^V$ is dominated by the spectrum analyzer and is calibrated on Au. Furthermore, our homemade STM is specially designed with sapphire to suppress the effect from any environment capacitance(*12*).

The experimental noise data is measured with bandwidth of 100 kHz or 400 kHz and the number of points for one frequency spectrum is 16384, 32768, or 65536 for different measurement systems. Throughout the paper, the shot noise power at each bias point is averaged for at least 10 minutes with 2000 or 4000 spectra.

## 4. Excluding other noise sources

We note that the enhancement of effective charge in shot noise measurements is very small (~0.01*e*). Consequently, other types of noise signal from the measurement system or the sample itself could introduce artifacts into the measurements.

To avoid such cases, we also measure effective charge on a normal metal Au(111), shown in the Extended Data Fig. 2. We can clearly see that $q = 1e$ for all the biases on Au(111) and $q > 1e$ when the bias $V < 40$ mV on the underdoped sample UD60K.

Additionally, we take the following steps to avoid artifacts.

First, we specially designed and carefully calibrated our amplifier system. Our noise was measured at around 4 MHz to suppress the contribution from low frequency 1/f noise(*13*) and random telegraph noise (RTN)(*10*). As mentioned above, we calibrated the background noise from different sources and also calibrated the gain to avoid small changes due to the environment.

Second, fluctuations in the measurement environment (such as temperature fluctuations) are compensated by measurement methods. During the measurement, a slow feedback is enabled to achieve optimal junction stability. Furthermore, we sweep the bias from a low value to a high value and then back to a low value many times and average the results at each bias level to minimize the effects of temperature drift over time and system instability. Extended Data Fig. 3 illustrates our method for averaging the results and defining the uncertainty of $q$. Some outliers are identified with detailed explanations provided in the Extended Data Fig. 3B-E.

Third, we consider the possibility of an unknown, additional noise source in our circuit (let us call it "mystery noise"). For example, shot noise from the leakage current between the source and gate of the HEMT (high electron mobility transistor) used in the amplifier could lead to an additional shot noise signal. Here we will show that we can exclude the possibility that our results are caused by a "mystery noise" source. First, our results show magnetic field dependence which excludes the noise sources without magnetic field dependence like equipment or local heating or tip gating effect. Next, it is not possible that our results are due to a magnetic field dependent current noise source for two reasons: Firstly, typical sources of extra noise, such as 1/f noise and RTN, are always voltage noise sources, not current noise sources. Secondly, any additional current noise source can be directly measured by our amplifier and will end up as a background noise by our calibration. Now, the only possible noise source left is a magnetic field dependent voltage noise source. If the noise source is independent of bias, for example additional noise from tip instability, it may lead to voltage dependent extra noise by $S_I = S_V/Z_{tot}(V)$, with $Z_{tot} = Z_{LC}//R_j^{diff}$. However, our amplifier is designed in a way in which $Z_{tot}(V)$ is dominated by $Z_{LC}$. Thus, a bias independent $Z_{LC}$ cannot lead to the bias dependent effective charge that we measured. Finally, a bias dependent voltage noise source like RTN can also be excluded because our amplifier works at 4 MHz, which suppresses the effects of low frequency noise like RTN. Additionally, RTN exhibits different behavior with bias(*14*) and it cannot show the strong position dependence behavior shown in Fig. 3. Furthermore, even if the possible "mystery noise" changes with different positions, it is unlikely to follow the same behavior as we calculated in Fig. 1F.

## 5. Sample preparation and STM measurements

The measurements in this work have been done in three different STM systems: a Unisoku USM1500 with an 8T magnetic field, a Unisoku USM1300 with a 9T magnetic field and a home-built ultra-stable STM(*12*), all of them equipped with MHz cryogenic amplifiers for shot-noise measurements and with base temperatures of 2.2K, 360mK and 4.2K, respectively. The magnetic field measurements were done in a Unisoku USM1500 for the UD58K sample and in a Unisoku USM1300 for the OP91K sample. The measurements without magnetic field on UD60K, UD70K and OP91K were done in our home-built STM.

The samples were cleaved in ultra high vacuum and at cryogenic temperatures (77K or 4K depending on the system) and immediately loaded in the STMs. Extended Data Fig. 4 shows a typical topographic image of a OP91K sample.

## 6. Extraction of $E_{pair}$ from shot noise measurements

The pairing onset energy, $E_{pair}$, discussed in the main text is extracted from the derivative of the effective charge, $dq/dV$. Extended Fig. 5 shows the simulation results using the same parameters from Fig. 1. At $V = \Delta_{PG}$, which is defined as the bias of the peak in the differential resistance, there is a dip in $dq/dV$, which is extracted as $E_{pair}$. Extended Fig. 6 shows an example of the extraction of $E_{pair}$ on sample UD60K. The energy of the dip in $dq/dV$ is indicated by gray solid lines and the full width at half maximum (red lines) is taken as the uncertainty of $E_{pair}$.

As we mentioned above, the amplifier is recalibrated using the shot noise measured at large bias voltages outside the gap. Different calibration ranges can affect the absolute value of $q$ extracted from shot noise. However, the value of $E_{pair}$ remains constant, as shown in the Extended Data Fig. 7

## 7. Extraction of $\Delta_{PG}$ and $\Delta_{kink}$ from the differential conductance measurements

The value of $\Delta_{PG}$ of individual spectra used in Fig. 3C is identified by the bias of the outer peaks in $dI/dV$. To extract this energy, we calculate the first derivative of the differential conductance and identify the energy at which it crosses zero.

For the phase diagram shown in Fig. 1A, spatially averaged values for different samples are used as $\Delta_{PG}$ and $\Delta_{kink}$. $\Delta_{PG}$ taken for each spectrum is obtained by fitting spectra with the extended Dynes formula(*15*). $\Delta_{kink}$ is identified by finding the point of inflection as the minimum in the second derivative of $dI/dV$, as shown in Extended Data Fig. 8. After extracting $\Delta_{PG}$ and $\Delta_{kink}$ for spectra at different positions, mean values are calculated and plotted in Fig. 1A.

## 8. Why shot noise in mesoscopic junctions does not reflect pairing.

We have recently shown that in typical mesoscopic junctions, pairing cannot be detected via shot noise measurements(*16*). Considering this, it is surprising that Zhou et al. (*17*) recently claimed to observe shot noise enhancement in La$_{2-x}$Sr$_x$CuO$_4$/La$_2$CuO$_4$/La$_{2-x}$Sr$_x$CuO$_4$ (LSCO/LCO/LSCO) junctions. Here, we show that this result is based on artifacts.

To determine whether the experimental setup of Zhou et al. (*17*) works as intended or generates artifacts, we consider their basic test measurements on commercial Nb-AlOx-Nb junctions. They report an enhancement of shot noise. However, looking carefully at their Extended Data Figure 8, one can see that the noise is enhanced at energies *outside* the spectroscopic gap. We have measured the same commercial Nb-AlO$_x$-Nb junctions, using an amplifier with less 1/*f* noise and less frequency spikes. Our data shows that there is no visible shot noise enhancement, in line with expectations from theory; thus, there is an artifact in Zhou et al. (*17*)'s setup that leads to apparent noise enhancement. We can only speculate on the reasons for this. First, contrary to their claims, there is significant 1/*f* noise in their bandwidth. Second, the equivalent circuit that they use to remove the effects of the capacitive shunting leaves out essential spurious effects, such as contact resistance. A further indication of problems in interpretation is that their data could only be explained by adding a virtual capacitor that changes in value with tunneling resistance, which is physically unreasonable. This effect could originate from random telegraph noise at the junction or contacts, which can lead to an enhanced noise in the measurements (note that random telegraph noise tends to be voltage dependent). The incorrect simplification of their circuit (Extended Data Fig.4 in their paper) can also lead to artifacts in the calculation of effective charge. One can see this in Niu et. al (*16*): in Fig 2, the voltage noise is seemingly enhanced when the differential tunneling resistance is large, and only when properly calibrated using the correct circuit component values one gets $q = 1e$ – using a wrong equivalent circuit would lead to an artifact. One can see in the test measurement of Zhou et.al (*17*) that they obtain such artifacts: when the differential conductance is enhanced, the noise is enhanced. Qualitatively, this is also what they see in LSCO/LCO/LSCO junctions, which are wrongly interpreted as charge bunching. We note that all these arguments also hold in the case of pinholes in the LSCO/LCO/LSCO junctions, however, given the high quality of the samples, they are not expected. We also note that the reported noise enhancement would be against expectations for any kind of superconducting gap, with or without multiple Andreev reflections.

# Supplementary Figures

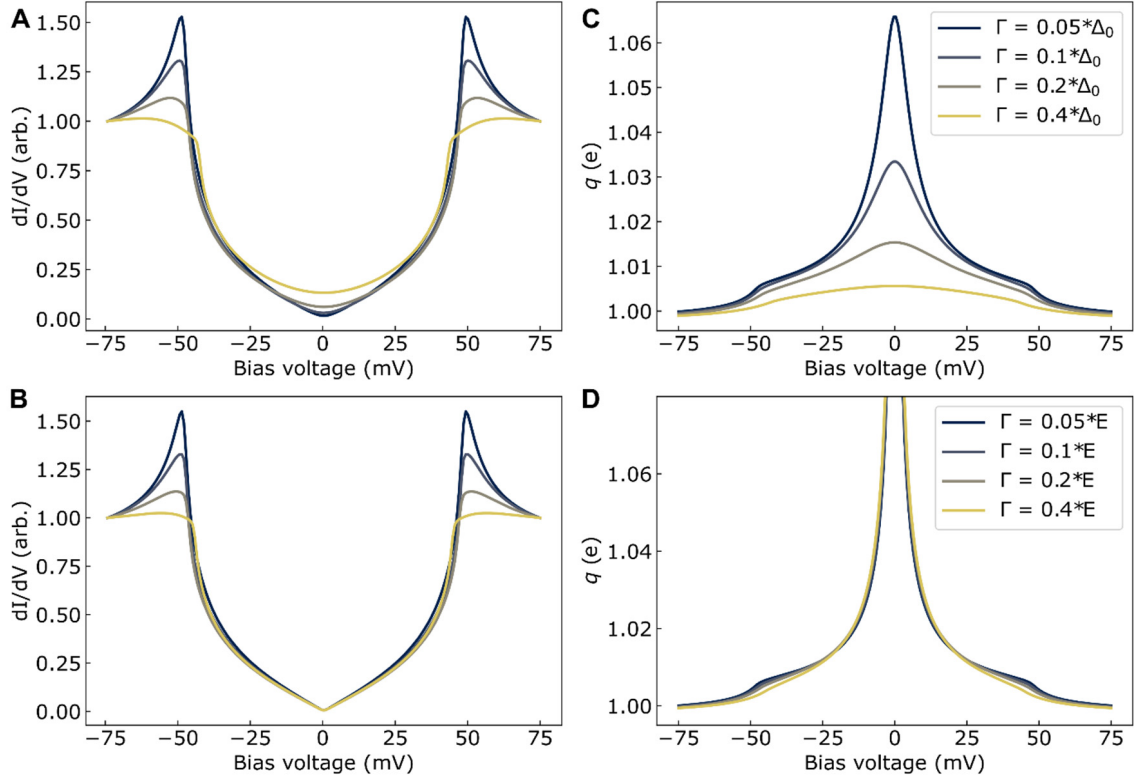

**Fig. S1. BTK simulation of a d-wave superconductor for different values of $\Gamma$.** (**A** and, **B**) Differential conductance and (**C**, and **D**) effective charge for different values of the inverse of the quasiparticle lifetime, $\Gamma$. **a** and **c** have a constant $\Gamma$ from darker to brighter: $0.05 * \Delta_0, 0.1 * \Delta_0, 0.2 * \Delta_0$, and $0.4 * \Delta_0$ (**B**) and (**D**) have an energy dependent quasiparticle lifetime: $\Gamma = \alpha * E$, where $\alpha$ is a constant that takes the values 0.05, 0.1, 0.2, and 0.4, from darker to brighter. The value of the maximum gap amplitude is $\Delta_0 =$ 52 meV for all calculations. The step feature in the effective charge becomes less obvious with larger $\Gamma$, making it more challenging for experiments in underdoped samples in which $\Gamma$ is typically larger. The temperature is $T = 4.2$ K and the transparency is $\tau = 2.5 \times 10^{-3}$.

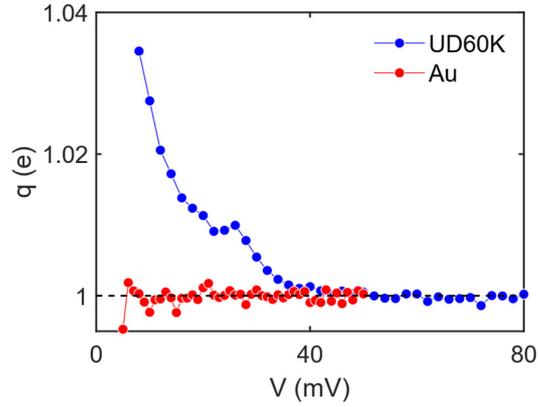

**Fig. S2. Effective charge measurement on sample UD60K and Au(111).** Blue and red dots are effective charge $q$ measured on UD60K and Au(111). On Au(111), $q = 1e$ are measured for all biases.

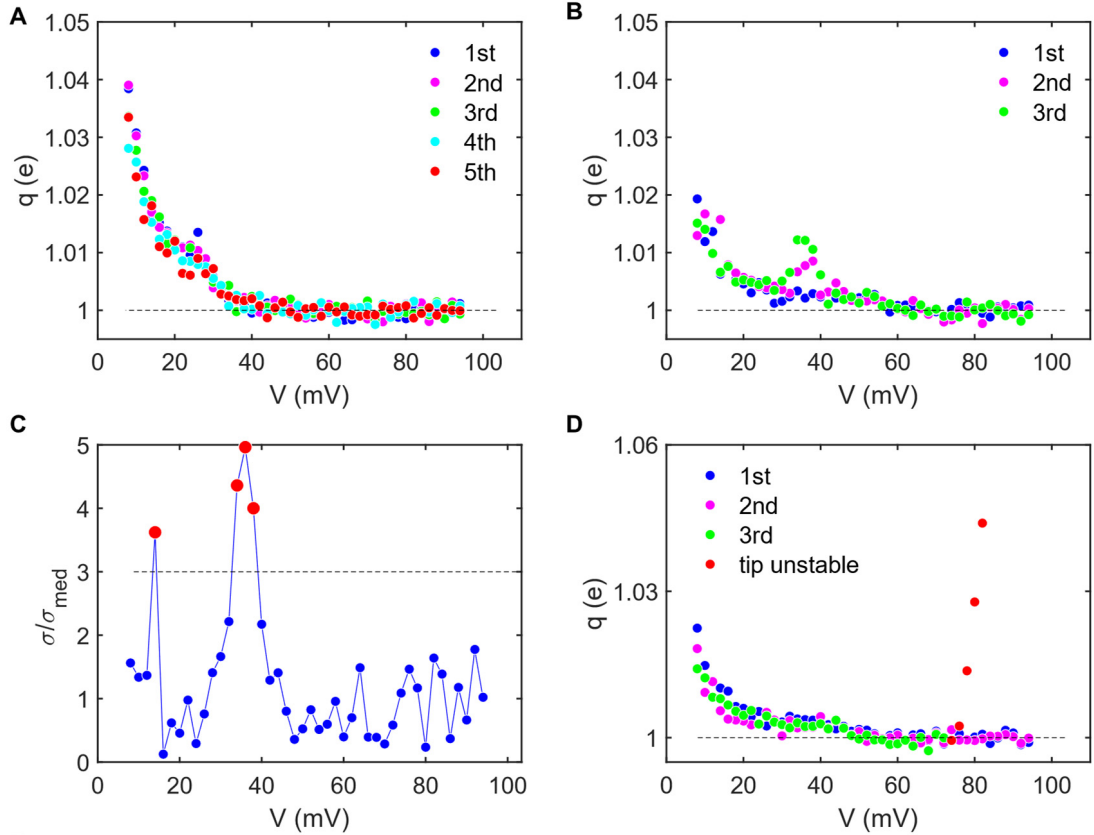

| | Number of positions measured | | | |
|---|---|---|---|---|
| | Total | no outliers | 0<outliers<5 | outliers>5 |
| UD60K | 6 | 4 | 1 | 1 |
| UD70K | 6 | 3 | 1 | 2 |
| OP91K | 6 | 4 | 1 | 1 |
| UD58K | Data are not averaged, outliers can not be defined | | | |

**Fig. S3. Uncertainty of $q$.** (**A**) One example of the results measured on UD60K. The effective charge $q$ at each bias is measured multiple times, as indicated by different colors. The average of $q$ at each bias is calculated and plotted in the main text. The standard deviation of $q$ at each bias is calculated as uncertainty and plotted as error bar in Fig. 3 of the main text. (**B**) One example of the results on UD60K with some outliers. Average and uncertainty of these results are plotted as purple dots in Fig. 3A of the main text. (**C**) Relative uncertainty of the results in (**B**). $\sigma = \varepsilon_S/S_I$ is the standard deviation of shot noise power $\varepsilon_S$ at each bias divided by shot noise power $S_I$ at the same bias. $\sigma_{med}$ is the median value of $\sigma$. The results with $\sigma/\sigma_{med} > 3$ (indicated by red dots) are defined as outliers. These outliers will not be used to calculate $E_{pair}$ as shown in Extended Data Fig. 4. There is one curve where the tip becomes unstable at the end of the measurements, as shown in (**D**). For these results, outliers (marked by red dots) are removed manually, and the rest of the results are used in calculations. (**E**)Summary of the results including the number outliers. For OP91K, results where $\sigma/\sigma_{med} > 7$ are considered as outliers because there are insufficient data points to accurately calculate $\sigma$.

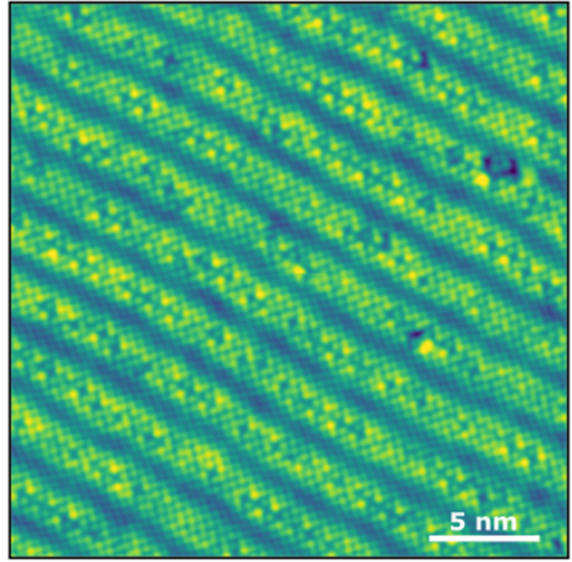

**Fig. S4. Topographic image of OP91K sample.** A typical 25 x 25 nm topographic image of $Bi_2Sr_2CaCu_2O_{8+\delta}$ measured on OP91K sample with set-up conditions: $V = 300$ mV and $I = 80$ pA. Crystal 'supermodulation' can be seen clearly.

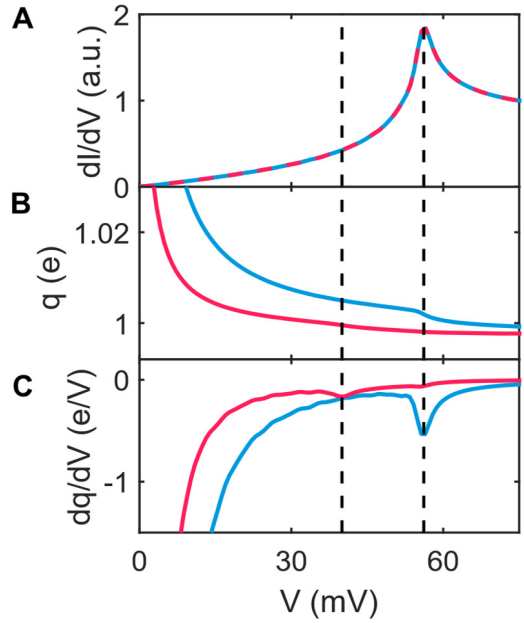

**Fig. S5. Extracting $E_{pair}$ from simulations.** Differential conductance $dI/dV$, effective charge $q$ and derivative of effective charge $dq/dV$ are calculated by BTK formulas. The same parameters as in Fig. 1f are used here. The dip in $dq/dV$ is identified as $E_{pair}$ and indicated by dashed lines.

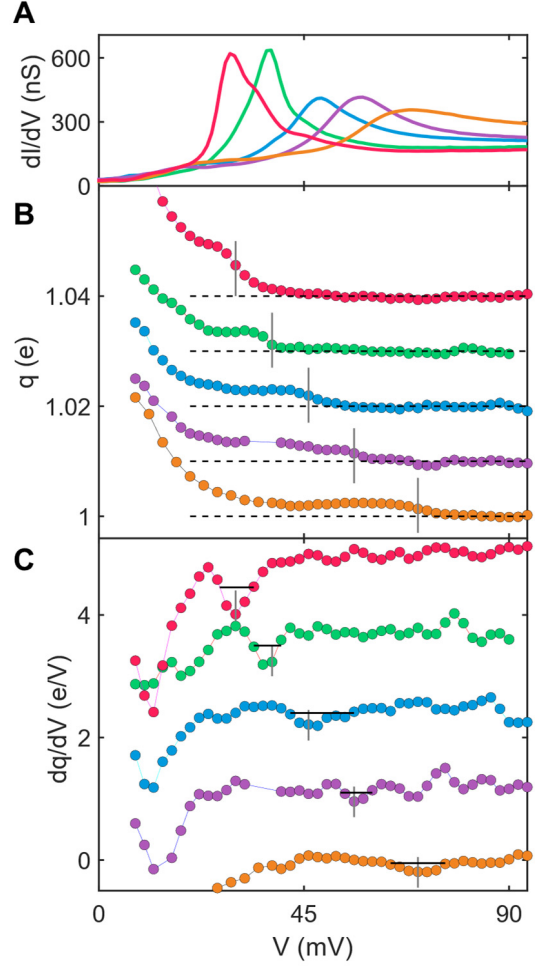

**Fig. S6. Extracting $E_{pair}$ and uncertainty of $E_{pair}$ for UD60K.** (**A**) Differential conductance for results at different positions. (**B**) Effective charge $q$ smoothed by a 3-point moving average. Outliers are hidden and not included in the smoothing process. More information on outliers is shown in Extended Data Fig. 3. (**C**), Gradient of $q$ smoothed for different positions. The dip is identified as $E_{pair}$ and marked by gray lines. The black horizontal lines show the full width at half maximum value which is taken as the uncertainty of $E_{pair}$ and plotted in Fig.3c of main text.

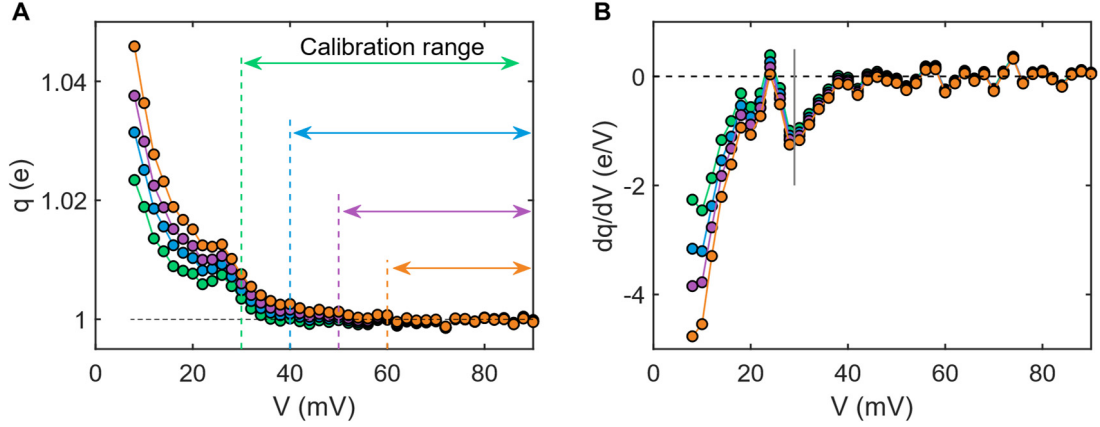

**Fig. S7. Absolute value of *q* varies with amplifier calibration range.** As mentioned in the methods, shot noise results outside $\Delta_{PG}$ are used to calibrate the effective gain of the amplifier to suppress the affect of gain drifting over time. The absolute value of *q* varies with different amplifier calibration ranges, but $E_{pair}$ remains unchanged. (**A**) An example of results on UD60K. Different colors correspond to different calibration ranges. The value of *q* is smaller with a larger calibration range because $q = 1$ is assumed in the calibration. (**B**) $dq/dV$ for different calibration ranges. $E_{pair}$ remains unchanged and is indicated by gray solid line.

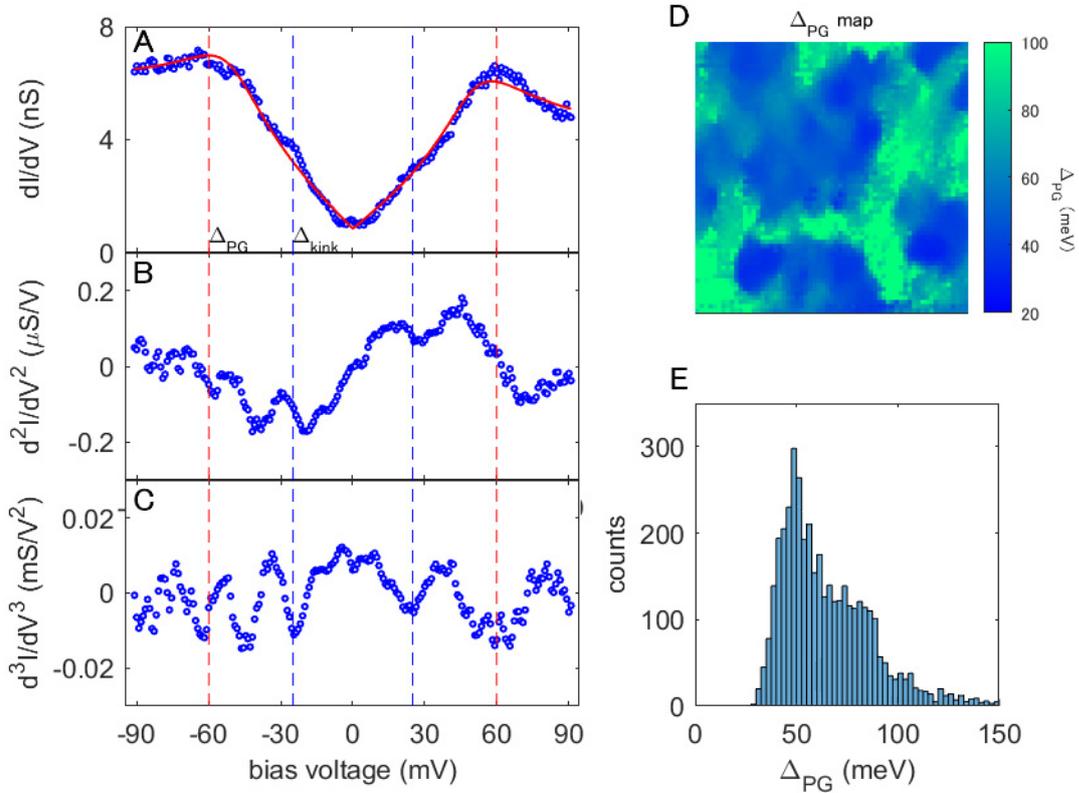

**Fig. S8. Extracting $\Delta_{PG}$ and $\Delta_{kink}$ from tunneling spectra map.** (**A**) One of the individual curves in a d*I*/d*V* map taken on UD58K at *T* = 2.2 K (blue dots) and the fitting curve (red

solid line). (**A** and **C**) Derivative and second derivative curves of d*I*/d*V*. Red and blue dashed lines indicate the $\Delta_{PG}$ and $\Delta_{kink}$, respectively. (**D**) $\Delta_{PG}$ map obtained by Dynes fittings(*15*) for all curves of the d*I*/d*V* map. (**E**), Histogram of (**D**). The mean value of all values of $\Delta_{PG}$ is plotted in the phase diagram shown in Fig. 1a of the main text.

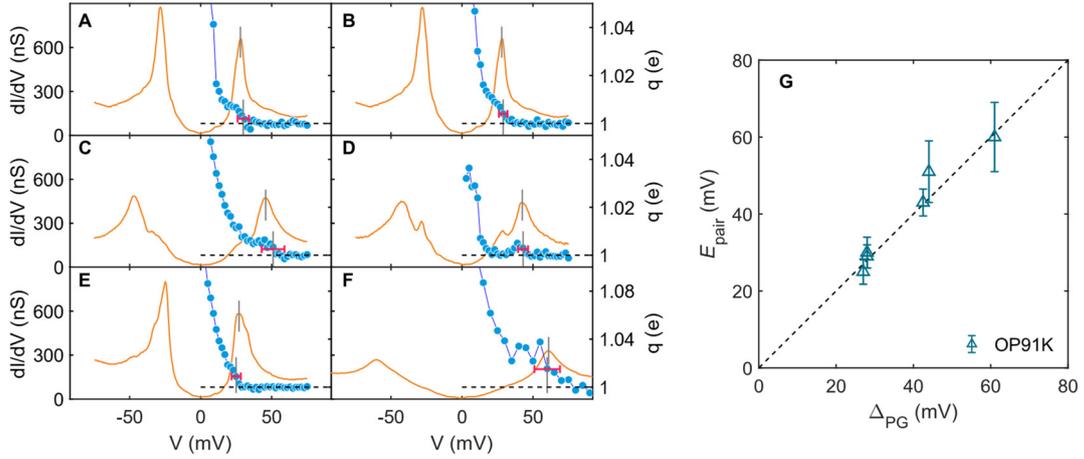

**Fig. S9. More results of OP91K at $B = 0$ T.** (**A** - **F**) Effective charge (blue dots) and $dI/dV$ spectrum (orange lines) at different positions of OP91K. $E_{pair}$ and $\Delta_{PG}$ are marked by gray lines. Uncertainties of $E_{pair}$ are shown as red error bars. (**A** - **E**), are measured from one sample at $T = 4$ K and **f** is from another sample at $T = 360$ mK. (**A** and **B**) are measured at the same position at different times and the average of these two results are plotted in Fig.3c of main text. (**G**), Summary of $E_{pair}$ vs. $\Delta_{PG}$ of OP91K samples.

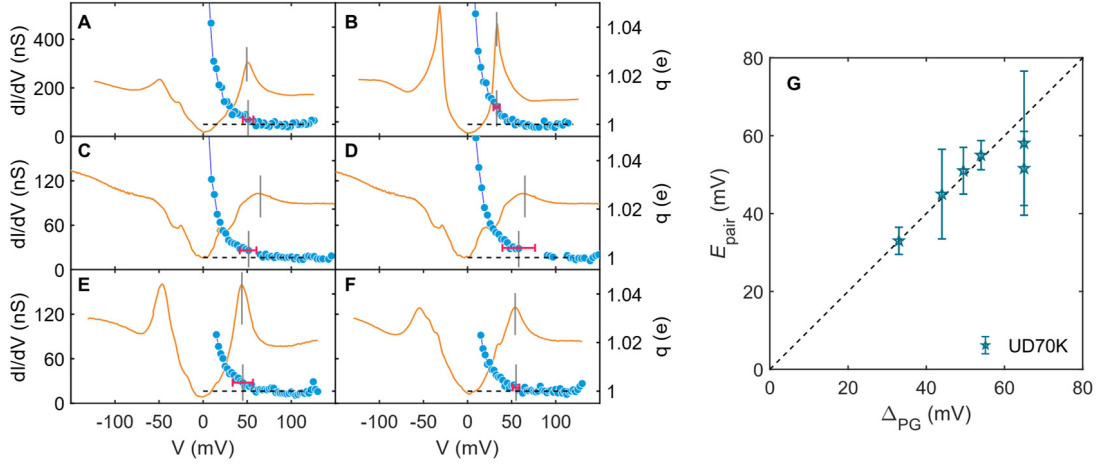

**Fig. S10. More results of UD70K at $B = 0$ T.** (**A** - **F**) Effective charge (blue dots) and $dI/dV$ spectrum (orange lines) at different positions of UD70K at $T = 4$ K. $E_{pair}$ and $\Delta_{PG}$ are marked by gray lines and uncertainty of $E_{pair}$ are shown by red error bars. (**C** and **D**) are measured in the same position. Some outliers with a standard deviation larger than three times the median value are not shown in the plot. (**G**), Summary of $E_{pair}$ vs. $\Delta_{PG}$.

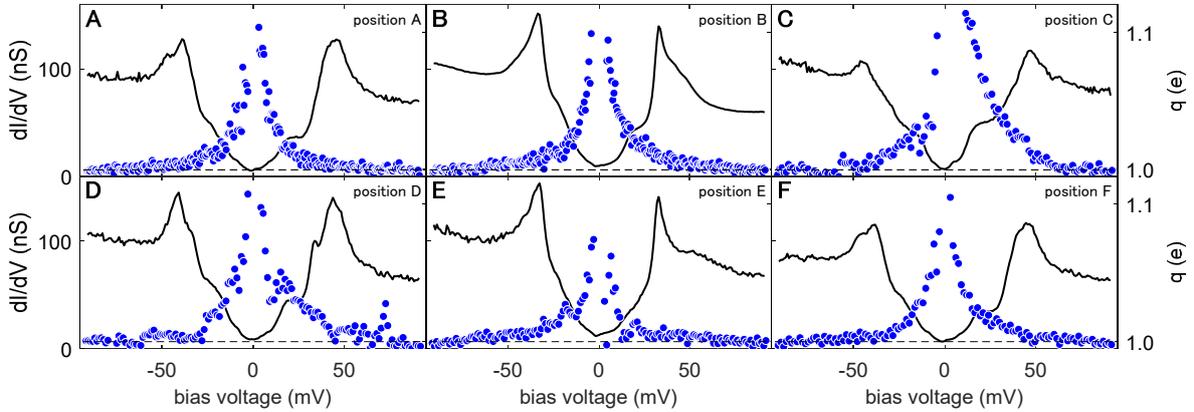

**Fig. S11. Results of UD58K at $B = 0$ T.** Effective charge (blue dots) and $dI/dV$ spectrum (black lines) at different positions. The values of effective charge for 0 T in Fig.4c of the main text are extracted from these data.

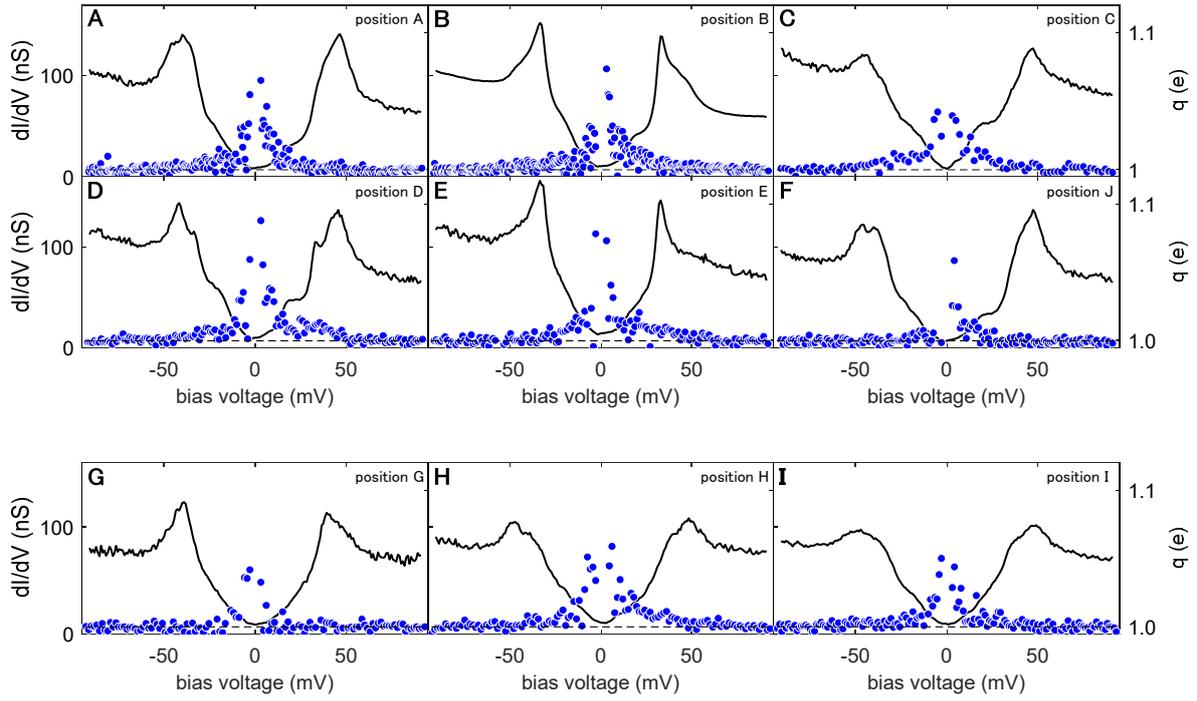

**Fig. S12. Results of UD58K under magnetic fields.** (**A** - **I**)Effective charge (blue dots) and *dI/dV* spectrum (black lines) are taken under $B = 1.0$ T (**A** - **F**)and $B = 1.4$ T (**G** - **I**),respectively. The values of effective charge under magnetic fields in Fig.4C of the main text are extracted from these data.

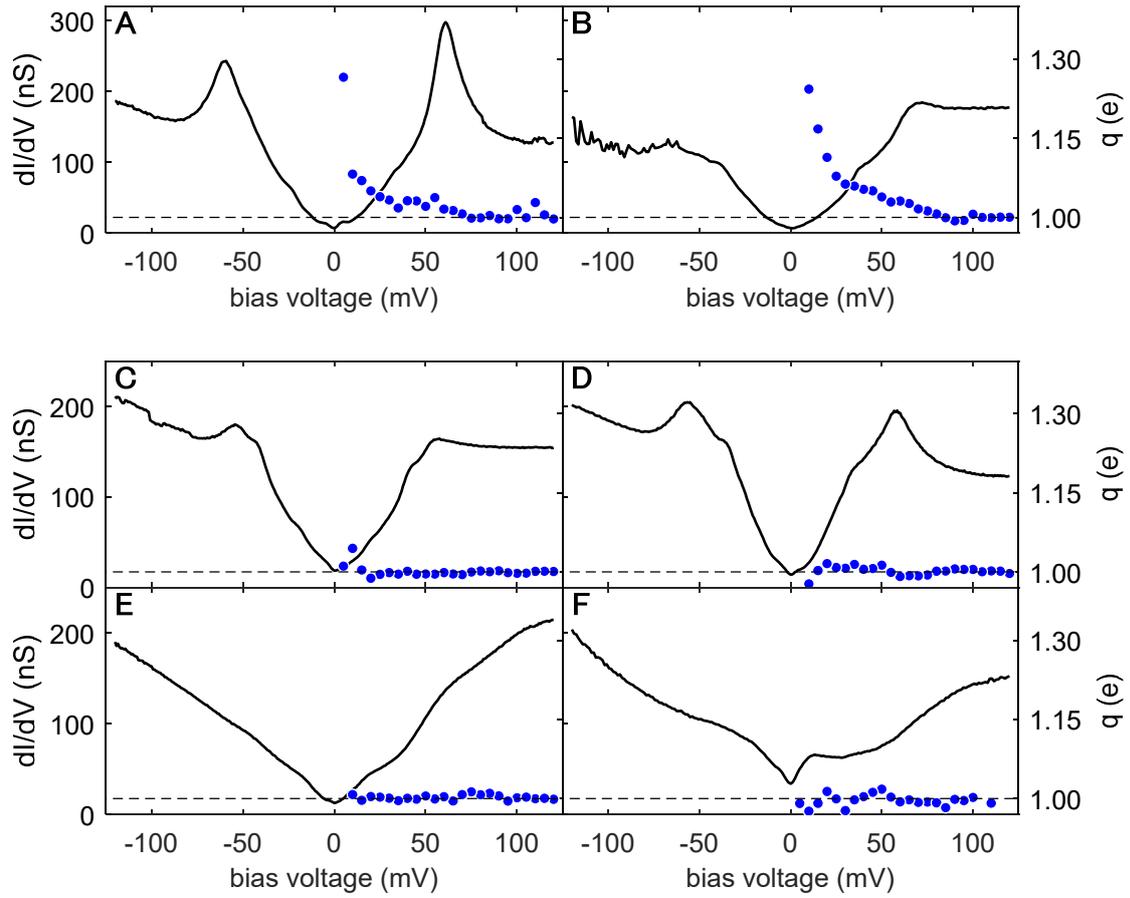

**Fig. S13 | Effective charge and dI/dV spectrum of OP91 K under magnetic fields.** (**A** - **F**) Effective charge (blue dots) and $dI/dV$ spectrum (black lines) are taken at $B = 0$ T (**A** and **B**) and $B = 6$ T (**C** - **F**). The values of the effective charge in Fig.4D are extracted from these data.

# Supplementary References